# Computer Vision-Based Health Monitoring of Mecklenburg Bridge Using 3D Digital Image Correlation

**Mehrdad S. Dizaji**, Graduate Research Assistant,

University of Virginia

Department of Engineering Systems and Environment

151 Engineer's Way, Charlottesville, VA, 22904-4742, USA

Phone:(434)987-9780, Email: ms4qg@virginia.edu

**Devin K. Harris**, Ph.D., Associate Professor,

University of Virginia

Department of Engineering Systems and Environment

151 Engineer's Way, Charlottesville, VA, 22904-4742, USA

Phone: (434) 924-6373; Fax: (434) 982-2951, Email: dharris@virginia.edu

**Bernie Kassner**, Ph.D., P.E., Research Scientist

Virginia Transportation Research Council

530 Edgemont Road, Charlottesville, VA 22903

Phone: (434) 293-1929, Email: bernie.kassner@vdot.virginia.gov

**Jeffrey C. Hill,** P.E., District Structure and Bridge Engineer

Virginia Department of Transportation - Richmond District

2430 Pine Forest Dr., Colonial Heights, VA 23834

Phone: (804) 524-6139, Email: jeff.hill@vdot.virginia.gov

**Abstract**

A collaborative investigation between the University of Virginia (UVA) and the Virginia Transportation Research Council was performed on the Mecklenburg Bridge (I-85 over Route 1 in Mecklenburg County). The research team aided the Virginia Department of Transportation - Richmond District in the characterization of the bridge behavior of one of the bridge beams that had been repaired due to a previous web buckling and crippling failure. The investigation focused on collecting full-field three-dimensional digital image correlation (3D-DIC) deformation measurements during the dropping sequence (removal of jacking to support beam on bearing/pier). Additionally, measurements were taken of the section prior to and after dropping using a handheld laser scanner to assess the potential of lateral deformation or out-of-plane buckling. Results from the study demonstrated that buckling of the tested beam did not occur, but did provided a series of approaches that can be used to evaluate the effectiveness of repaired steel beam ends. Specifically, the results provided an approach that could estimate the dead load distribution through back-calculation.

1. Introduction

Highway bridges represent some of the most critical components of the transportation network, providing pathways to traverse obstacles (i.e. waterways, roadways, gorges, etc.) that interrupt a roadway corridor. This criticality applies to most bridges because they often serve as the most efficient pathway over these obstacles for many miles, but this criticality is most apparent on the U.S. interstate system, which supports both inter- and intra-state traffic at higher volumes than any other roadway type in the country. The origins of the national interstate system date back to the 1950's following the conclusion of World War II, but these routes and those developed since have evolved beyond their original intentions into the primary route for movements of people, goods, and services essential to the maintenance of the national economy.

In recognition that much of the transportation infrastructure was built during an era of national need and with a finite intended service life, a fundamental challenge that arises is derived from the trade-offs between performance, maintenance, and preservation and the user and economic impacts associated with interrupting service (i.e. traffic disruptions, lane closures, detours, etc.). According to modern design practices, these structural components are expected to endure relatively long service lives (50-75 years), but many of the interstate structures in service today are approaching or exceeding these original design service life intervals. For interstate bridges, the trade-offs between performance and maintenance/preservation practices often trend towards minimizing traffic disruptions except under the most extreme conditions. For many interstate bridges this results in maintenance and repair activities without complete bridge closure, requiring operations to be performed under live traffic [1-3].

The damage common to bridges has been well documented in the literature, but some common types of damage that occur in bridges include crack, corrosion, spalling, all of which could lead to structural failures if unchecked. Steel corrosion is a particularly common challenge in locations where the bridges are exposed to salt, whether from agency applied salt or salt from a marine environment exposure. In states where roads are salted to reduce freezing/icing on the surface, the salt eventually finds a pathway to elements or components that are vulnerable to corrosion, such as internal reinforcement in concrete and steel beam ends. For the latter case, this exposure can eventually result in section loss, which typically occurs in the base of the beam web, the beam bottom flange, and bottom web/flange intersection. If this section loss is allowed to progress, the bearing capacity in this region could be reduced to the point that web crippling or shear yielding failures could result from the change in geometry resulting from the deterioration. In most instances interventions are often initiated by the owner agencies prior to failure, which include interventions such as stiffening with supplemental plates or partial section replacement. For the partial section replacements, part of the beam is removed and replaced with a new section or partial section of similar geometry. Following intervention, the bridge is placed back in service. These repairs have been successfully performed for years within the bridge industry; however, very little information is available regarding the performance of these repairs.

2. **Motivation**

In 2017, safety inspectors in the Commonwealth of Virginia observed that some beams in one of the interstate bridges experienced buckling in the web and the bearing stiffeners. There was also some deterioration in the bottom flange. The bridge (Fed ID 11920, Virginia Structure No. 058-2008) carried northbound Interstate 85 over U.S. Route 1 in Mecklenburg County, Virginia.

Herein the bridge is referred to as the Mecklenburg Bridge. Built in 1965, the structure averaged 10,374 vehicles per day, with 19% of those being trucks. The structure has four simple spans, measuring 77.5 ft., 105.5 ft., 91.5 ft., and 86 ft. (for a total of 360.5 ft.) with a 51° skew. The superstructure consisted of six steel beams spaced at 8 ft., 7⅛ in, supporting an 8-in composite concrete deck and braced by intermediate diaphragms.

In 2018, contractors began repairing the buckled sections noted during the 2017 safety inspection. The span in question was the 91.5-ft span (Span 3), where the beams consisted of built-up sections, with 60-in by ⅜-in steel plate for the web, 12-in of steel plate of varying thicknesses for the top and bottom flanges. The bearing stiffeners were 5¾-in by ½-in plates, while the intermediate stiffeners were 4½-in by ⅜-in plates. Diaphragms were spaced at 4 ft., $10^{15}/_{16}$ in. On the exterior face of the girder, two 4½-in by ⅜-in plates served as horizontal stiffeners along the length of the beams. Originally, the beams were set on steel rocker plates on the expansion ends. However, these plates were replaced by laminated elastomeric bearing pads in 2000. These pads had an unusual shape in plan and were placed at an obtuse angle relative to the centerline of the beams. The repair began by jacking the beams up off of the bearings, cutting out the buckled sections and welding replacement steel into place. The contractors also increased the thickness of the bearing stiffeners and added some additional transverse stiffeners near the bearing but closer to mid-span. However, four days after setting the repaired girder back on the bearing, inspectors from Virginia Department of Transportation (VDOT) noticed that the end of the web had buckled again. There was no buckling in the stiffeners. However, there was uncertainty as to when, exactly, the buckling occurred, and whether the cause was simply vehicular loads, dead loads, or otherwise. Additionally, there was some uncertainty as to whether the elastomeric pads were causing additional stress at the end of the beam. In response to this buckling issue, the Virginia Department of Transportation's Richmond District coordinated a collaborative investigation between the University of Virginia and the Virginia Transportation Research Council (UVA/VTRC) to evaluate the cause. This evaluation focused on monitoring a specific girder (i.e., the buckled girder which was repaired by VDOT) as the contractors set the member back on its bearing, and limited continuous monitoring of the girder for several days after setting. This following sections describe the experimental program and findings from this investigation.

## 3. Evaluation Approach

Historical practice for evaluating the performance of in-service structures has typically utilized deformation measurement tools such as strain gauges, linear variable displacement transducers and vibrating wire gauges [4-6]; however, for a forensic analysis these tools are limited to local measurements and not suitable for characterizing the response of a surface. Recent advances in photogrammetric and image-based measurement techniques [7-9] provide an alternative strategy for describing structural response and characterizing deformations [10]. Additionally, these approaches are attractive for civil infrastructure systems due to their non-contact nature, relative ease of deployment, and recent improvement of imaging technologies. To evaluate the cause of the unexpected buckling following the post-repair buckling on the Mecklenburg Bridge, the UVA/VTRC collaboration utilized three-dimensional digital image correlation (3D-DIC) measurements of the mechanical response during the lowering process, three-dimensional scanning (3DS) before and after lowering, and localized continuous strain monitoring at specified locations. The following subsections provide an overview of the 3D-DIC and 3DS technologies used in this study.

### 3.1. Three-dimensional Digital Image Correlation

Digital image correlation (DIC), describes a non-contact photogrammetric technique capable of measuring the full-field deformation response of a structural component under loading. DIC relies on the general concept of tracking the movement and deformation of pixel patterns over a sequence of digital images. The measured component is typically patterned to create a high-contrast speckled surface, with the correlation or pattern matching algorithm constrained to a subset region to ensure correspondence of unique patterns [11-17]. Across the sequence of images, deformations are measured by tracking pixel movements through space with interpolation used to describe the full-field response across the specimen surface. Two-dimensional digital image correlation (2D-DIC) leverages planar image sets and can be used to measure in-plane deformation, whereas three-dimensional digital image correlation (3D-DIC) extends the principles of photogrammetry using calibrated stereo-paired cameras to measure three-dimensional deformation [18-23]. The ability to measure deformation have always been vital to the field of experimental mechanics, where DIC was discovered, but recent advances in imaging tools and the availability of reliable commercial DIC tools has created opportunities for extending the technique to large scale infrastructure applications. The DIC system used in this work was from Correlated Solutions Inc. (CSI), and included the software packages VIC-Snap for data acquisition and VIC-3D for post-processing [24]. A comprehensive treatment of DIC is available in the literature [11-23] and not presented here, but additional details on the DIC deployment used in this investigation are provided in a later section.

## 3.2. Three-dimensional Scanning

Three-dimensional scanning (3DS) characterizes measurement techniques capable of capturing the three-dimensional shape of real-work objects/spaces and translating these measurements into digital representations. 3DS includes systems that rely on white or structured light capture, laser scanning, and stereo-photogrammetry. Each of these techniques requires line-of-sight for measurement, with resolution of the measurement typically determined by the selected instrument, which includes a spectrum of systems ranging from tablet-based systems to high-resolution terrestrial and handheld systems. Regardless of the system type, each is able to generate three-dimensional point cloud representations of the measured surface via time-of-flight measurements or geometric reconstruction approaches. The system used in this work, HandySCAN 3D 700, was manufactured by Creaform and relies on the measurement principles of stereo-photogrammetry for measuring surface properties of the object. This system uses a series of laser lines to illuminate the surface of the object being measured, with the surface topology deviation captured in real-time using stereo-paired cameras. Positioning targets on and around the measured object allow for dynamic self-positioning within 3D space. Manufacturer specifications on the system are presented in Table 1, with additional details on the performance of this system available in [25].

*Table 1: Manufacturer Specifications for Creaform HandySCAN 3D 700 [25]*

| | |
|---|---|
| Accuracy | Up to 0.030 mm |
| Resolution | 0.050 mm |
| Measurement rate | 480k measurements/s |
| Light source | 7 laser crosses (+1 extra line) |
| Scanning area | 275 x 250 mm |
| Stand-off distance | 300 mm |
| Depth of field | 250 mm |

## 4. Experimental Study

The experimental investigation focused on the evaluation of a single beam end location following repair (Figure 1). This evaluation included the measurement of the system response during the drop sequence and a subsequent evaluation after the structure was placed back in service. The following subsections describe the 3D-DIC and 3DS experimental configurations.

### 4.1. Image-based Diagnostics (3D-DIC)

In this work, full-field 3D-DIC measurements were acquired only during the dropping phase of the repair, which included the transition from unloaded to supporting bridge dead loads over a time interval of approximately 1 minute. The 3D-DIC measurements provided the ability to characterize the deformation response in three-dimensions within the field of view of the stereo-paired cameras used. During this testing phase, 3D-DIC measurements were collected on two regions on opposite sides of the beam end: 1) Region 0 (System 0) evaluated the web panel region beyond the vertical bearing stiffener and 2) Region 1 (System 1) evaluated the tail end of the beam that extended beyond the bearing support. Two separate regions were collected because parts of the beam end were obscured from the camera depending on the side of measurement.

For the DIC measurements, the workflow consisted of specimen preparation, camera setup (focusing, calibration, and image acquisition), and post-processing of results. For the specimen preparation, the surface of the beam within the area of interest was cleaned, painted with a flat white base coated and covered with a dense and random speckle pattern using permanent markers for the correlation process. The camera setup included the use of two Point Grey Grasshopper 5-megapixel charge coupled device (CCD) cameras (resolution 2448 x 2048 @ 7.5 fps) with fixed lens configurations, 12 mm for System 0 and 12 mm for System 1. The camera pairs were positioned ~3 feet from the measurement surface which created a field of view of ~ 2.5 x 2.5 feet. Using the image acquisition package (VicSnap), the camera calibration was performed to determine intrinsic and extrinsic camera parameters. Following calibration, the data acquisition consisted of image collection (@ 2 fps) from both camera pairs during the dropping procedure. During post-processing, the area of interest was split into rectangular windows or "subsets" and the correlation algorithm tracked the unique patterns of speckles within each subset across subsequent frames. The patterns in the subsets were tracked on a grid of a specific "step" size, which dictates the spatial resolution of the calculated points. To achieve a fine grid of unique patterns in subsets, the selection of the subset size was achieved through direct experimentation and a square subset of 35 pixels at a step of 6 pixels was selected (Figure 1a-b). Additional details on DIC are available within the literature [11-23].

In addition to the DIC measurements, foil resistance gauges (1/2" 120Ω) were installed at select locations with appropriate weatherproofing for a long-term monitoring program beyond the scope of this work. However, it should be noted that the locations for the strain gauges and wiring interfered with the line of sight of the DIC measurements and data is not available in this small region. The results of these long-term strain gauge sensor measurements are not described in detail in this work, but the sensors were evaluated during preliminary laboratory testing and successfully demonstrated correspondence between strain measurements (DIC and resistance-based) techniques and briefly described below.

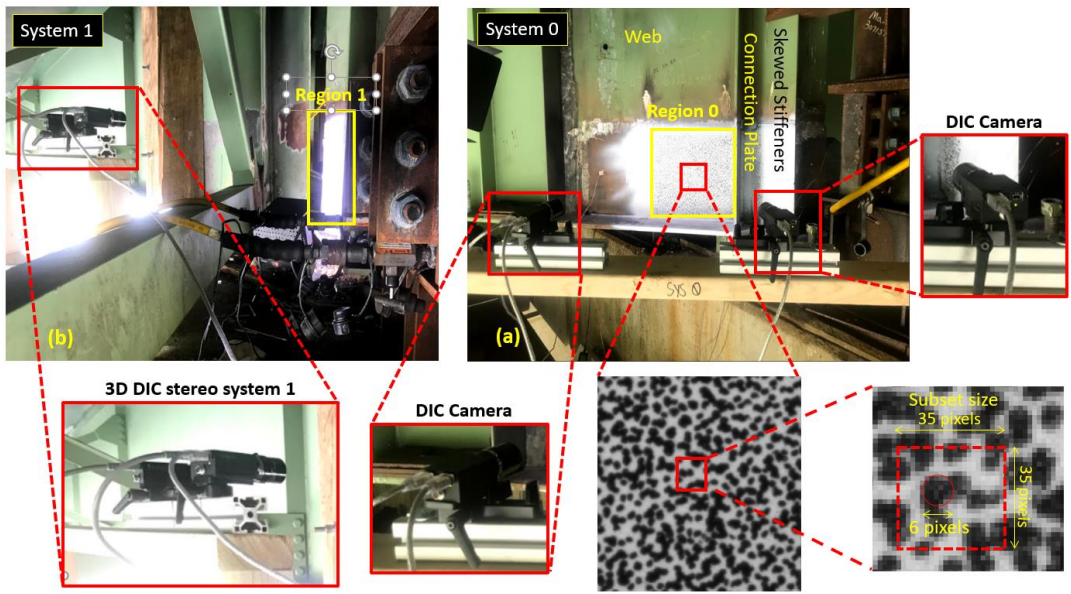

(a)

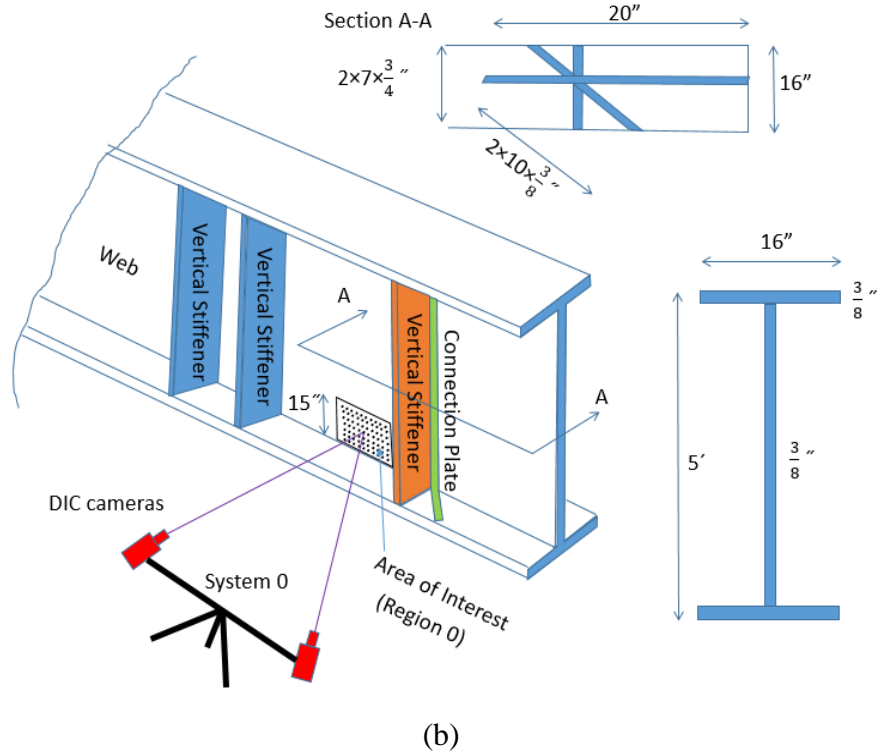

(b)

*Figure 1. Experimental DIC setups for the Region 0 and 1, prior to dropping the beam, (a) Region 0 (Web Panel, System 0), (b) Region 1 (Tail End Region, System 1), (b) Schematic of beam geometry for Region 0 (Note: Region 1 located beyond the web which is not visible from this perspective*

### 4.1.1. Laboratory Simulation and Validation

To evaluate the correspondence between the measurements derived from traditional foil gauges and those from 3D-DIC, a limited laboratory study was performed. This study, followed the framework form previous works of the research team [26-29] and utilized a A992 W 10x22 steel beam patterned on the web for DIC and instrumented at the top and bottom of the web with the same foil resistance gauges used in the field study. the beam was selected based on the structural laboratory limitations such as location of MTS actuators and the loading frames. The steel beam was tested within the structures laboratory at the University of Virginia in a bearing configuration with the beam supported by a rigid pedestal and loaded to 8 kips using a MTS 35-kip actuator in displacement-control over a series of compression cycles (2 Hz) to deform the beam within the elastic range. Similar to the field measurements, DIC and foil gauge measurements could not be collected at the exact same location due to loss of data in the regions of the sensor; however, DIC data was collected just above and below the gauge locations (Figure 2). Results from the sensors and DIC measurements are presented in Figure 2 and clearly demonstrate the consistent results derived from both measurement approaches.

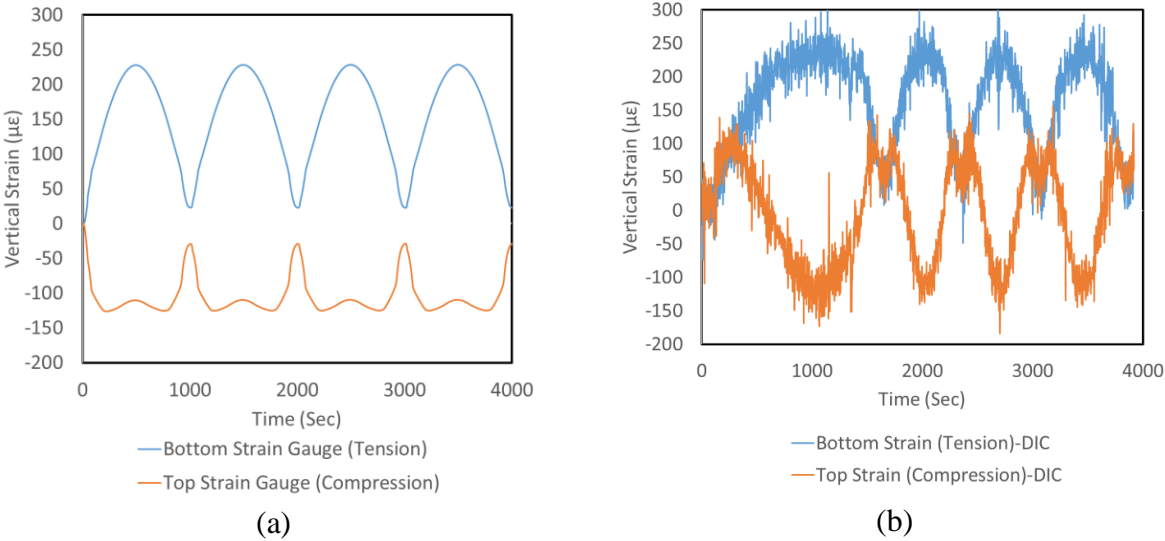

(a)          (b)

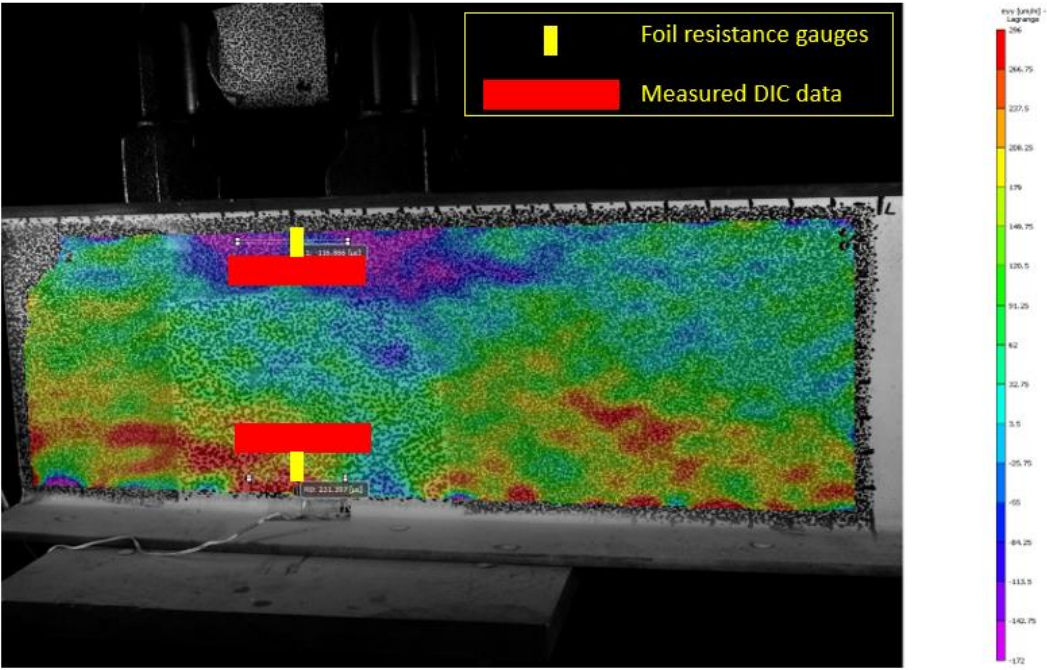

(c)

*Figure 2. The lab simulation results for the regions, (a) foil resistance gauge results (b) 3D DIC results at the corresponding location of the foil gauges, (c) Vertical full field DIC strain results from sinusoidal loading (at peak load of 8 kips)*

**4.2. Three-Dimensional Scanner-based Diagnostics**

In addition to the 3D-DIC measurements acquired during the dropping sequence, a series of 3DS measurements were taken before the start of the drop sequence (Scan 1) and two weeks after the dropping sequence (Scan 2). The goal of the scanning measurement was to evaluate whether any out-of-plane deformation occurred within the web panel section (Region 0). It should be noted that Region 2 was not scanned due to the difficulty of maneuvering the scanner within the tight space surrounding the tail section, which was also influenced by scanner operational constraints on the distance between the scan surface and the imaging sensors. Retroreflective markers (illustrated by the white/black markers) were applied to the surface prior to the 3D-DIC speckle patterning and then removed and then reapplied prior to Scan 2. It should also be noted that the foil gauges also had implications in the measurements from the 3DS, which are described in a later section.

## 5. Results and discussions

Based on the outcomes of previous repairs, it was expected that some form of local buckling would occur following the dropping process; however, no significant buckling was observed during the period of evaluation (during dropping or during follow-up scanning). While buckling was not observed during the period of performance, the results from the 3D-DIC measurements were used to estimate the effectiveness of the repair solution.

**5.1. 3D-DIC Results**

Measurements derived from DIC were able to describe the full-field 3D deformation response over the loading. Figure 3a illustrates the full-field response of Regions 0 at the end of the drop sequence. With the foil strain gauge mounted on the observed surface, measurements were not available within these regions; however, the full-field measurement allows for evaluation of global and local responses that cannot be realized with a strain gauge. For example, Figure 3a represents the out-of-plane deformation of Region 0, which exhibits positive curvature with an amplitude on the order of approximately 0.00197 in. (~50 microns) along the diagonal. A similar observation can be made for the tail of the beam (Region 1) which exhibits a similar contour change along the diagonal; however, the magnitude of the deformation is significantly lower, on the order of 3-5 microns, suggesting no significant out-of-plane deformation in this region.

To evaluate the effectiveness of the repair, virtual strain gauges within the DIC post-processing were used to determine the response during the loading sequence. However, as previously noted, these gauges could not be aligned at the same locations as the foil gauges due to the lack of data in these regions. Virtual gauges did provide an illustration of the response of the system and a mechanism for comparison with the foil strain gauge measurements at similar locations. Figure 3b-d illustrates the vertical strain response at the three locations within the web section at time 100 seconds for Region 0, an average vertical strain along the stiffener in Region 0, and three locations within the web section for Region 1. A common characteristic observed amongst all of the measurements is the deformation in compression as the specimen is lowered onto the bearing pads. One exception is the foil strain gauge located approximately 12 inches from the bearing stiffener, which experiences a tensile response. The mechanism behind this unexpected deviation is unclear, but it should be noted that local tensile responses are also present within the DIC contour field (Figure 3b-d) even though the average strain across the virtual strain gauge regions consistently measure compressive strains. When comparing results derived from the foil strain gauges and those from the DIC measurements, it is clear that the results derived from DIC are consistently larger than those from the foil gauges; however, the exact cause is unknown and the previously described correspondence between the two in the laboratory study proved consistent, suggesting an anomaly within the foil gauge measurement. To evaluate the validity of the results, a rough estimate of the expected strain magnitude was determined to be ~ 200 $\mu\epsilon$, using an approximated load (i.e. beam self-weight, effective slab area) distributed over the bearing area (i.e. web and stiffeners) to determine stress, with the assumption of elastic behavior ($E_s$ = 29,000 ksi). While only approximate, this order magnitude aligns with those measuring using DIC. Also, time history of the results for the mentioned locations for the Regions 0 and 1 are plotted in Figure 4. As it can be seen in Figure 4, during dropping the beam, vertical strains increase suddenly and then after some time remained stable, which can be reasoned as the fact that the beam is settled and there is no more vertical deformation.

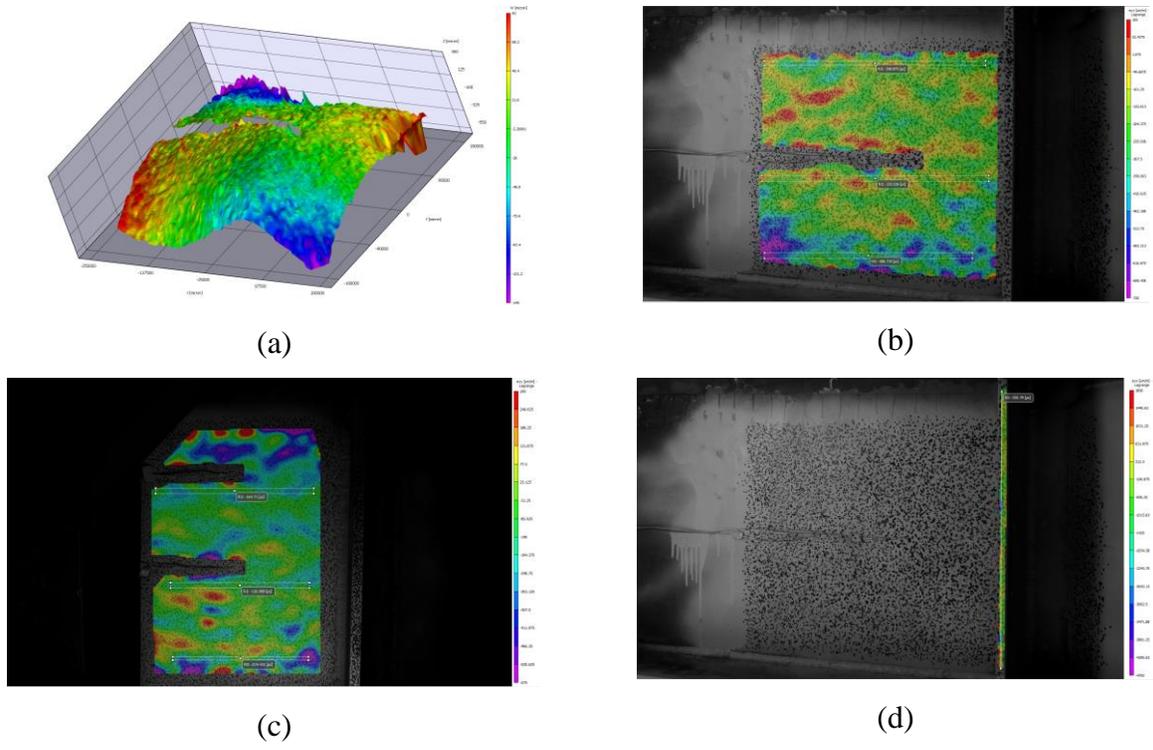

(a) (b)

(c) (d)

*Figure 3. (a) 3D Out-of-plane deflection (w) at t = 100s (Note 3 regions of interest) for Region 0. (b) Vertical strain measurements on the Region 0 (c) Vertical strain measurements on the Region 1, (d) Vertical strain measurements on the Connection Plate*

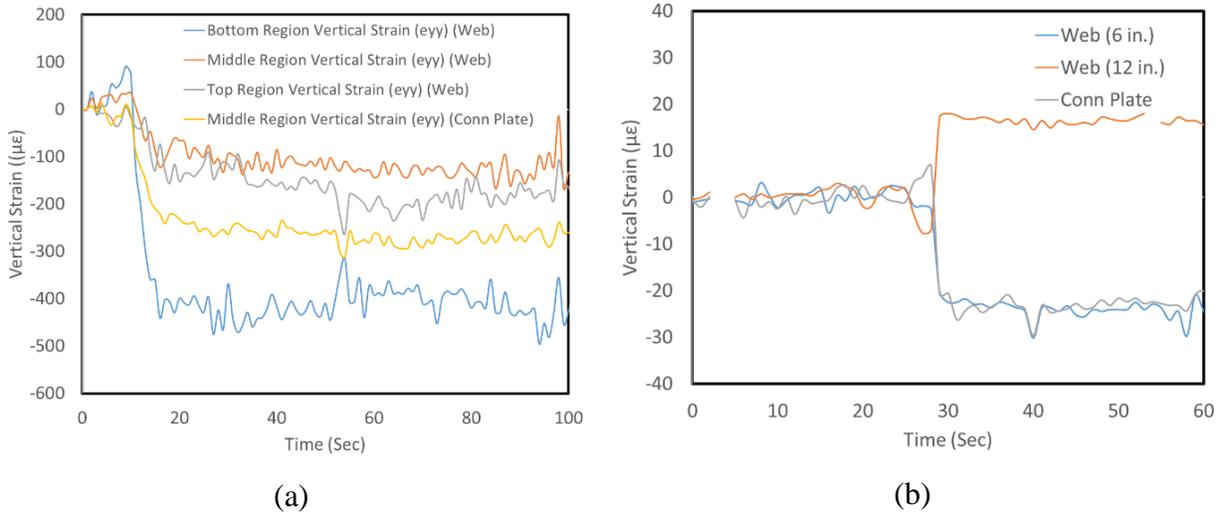

(a) (b)

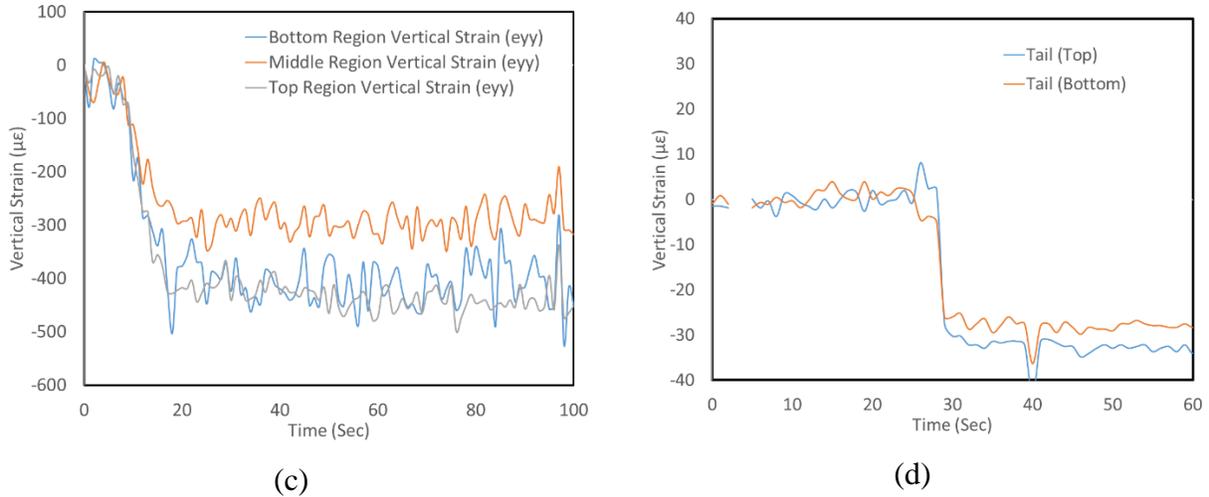

*Figure 4 – Vertical Strain/deformation for Region 0 and 1 during dropping (a) vertical strain ($\varepsilon_{yy}$) time history (System 0) from 3D DIC, (b) vertical strain ($\varepsilon_{yy}$) time history (System 0) from foil gauges, (c) vertical strain ($\varepsilon_{yy}$) time history (System 1) from 3D DIC, (d) vertical strain ($\varepsilon_{yy}$) time history (System 1) from foil gauges*

*Finite Element Analysis* - For further evaluation of the measured results, a simplified finite element model of the beam was developed in ABAQUS, a robust commercially available finite element software package [30]. The steel beam was simulated using a total of 5,500 Continuum 3D hexahedral solid elements (C3D8) with full integration, with the geometry of the beam developed using the bridge plans. For the boundary conditions, the supports were assumed to be simple with the additional dead loads applied as a distributed loading along the beam length in addition to the beam self-weight. The model was analyzed as a static elastic analysis to determine the strain distribution within the system. Figure 5 provides an illustration of the vertical strain for end region of the model with relevant locations highlighted. The results do not exactly agree with those derived from the DIC measurement at corresponding locations due to model approximations and uncertainty in boundary conditions, but the model results do reinforce previous findings related to appropriate order of magnitude for strain within the selected regions. It should also be noted that the idealized model was created without any initial imperfections and was therefore not expected to experience out-of-plane deformations due to symmetry of the model.

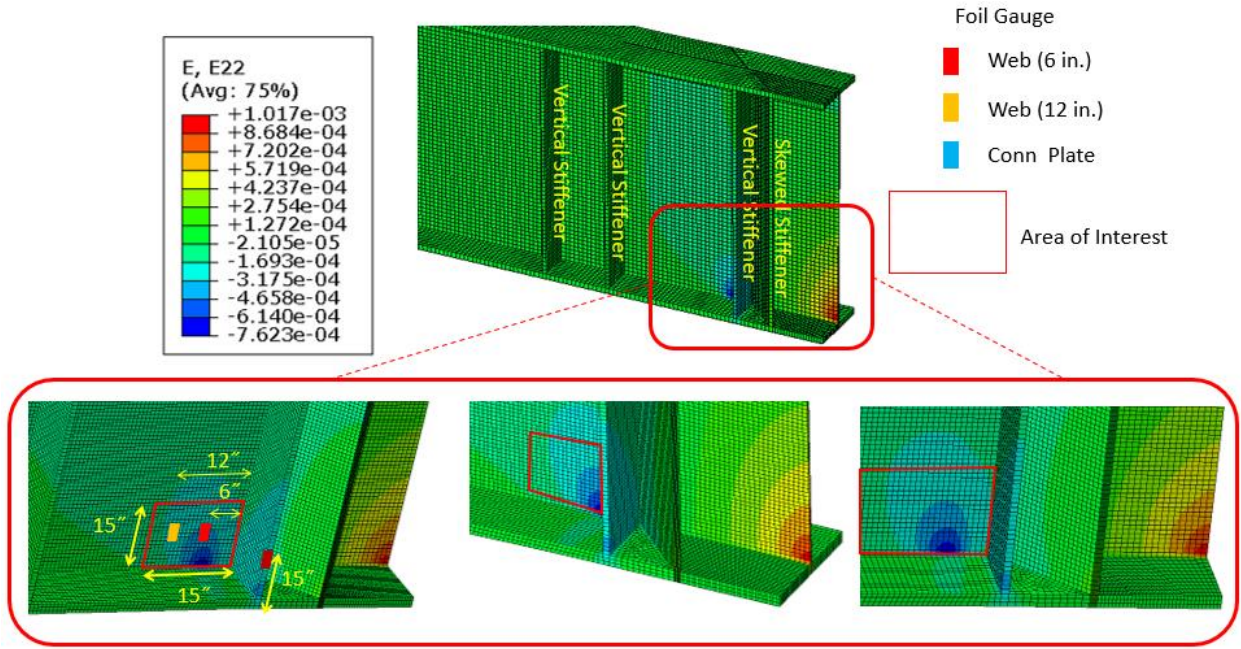

*Figure 5. Finite Element Model of the corresponding beam*

## 5.2. 3DS results

Similar to the 3D-DIC results, measurements derived from the 3DS were able to illustrate the full three-dimensional surface geometry of the beam end for Region 0, with the 3D shape reconstruction for each of the scans collected shown in Figure 6. In Figure 6b, it becomes apparent that deviations associated with the strain gauge weatherproofing and wires are captured within the second scan; however, no additional deformations are obvious to the naked eye. Following the surface reconstruction, the geometry of the two surface scans was aligned using common reference points between each scan and a deviation analysis was performed. The deviation analysis was performed using the Cloud Compare software package, which is an open source 3D point cloud and mesh processing software designed for direct comparison of 3D point clouds or 3D entities [31]. The deviation analysis confirmed the subjective observations from the 3D shape reconstructions, with only the deviations from the strain gauge systems highlighted in Figure 6c.

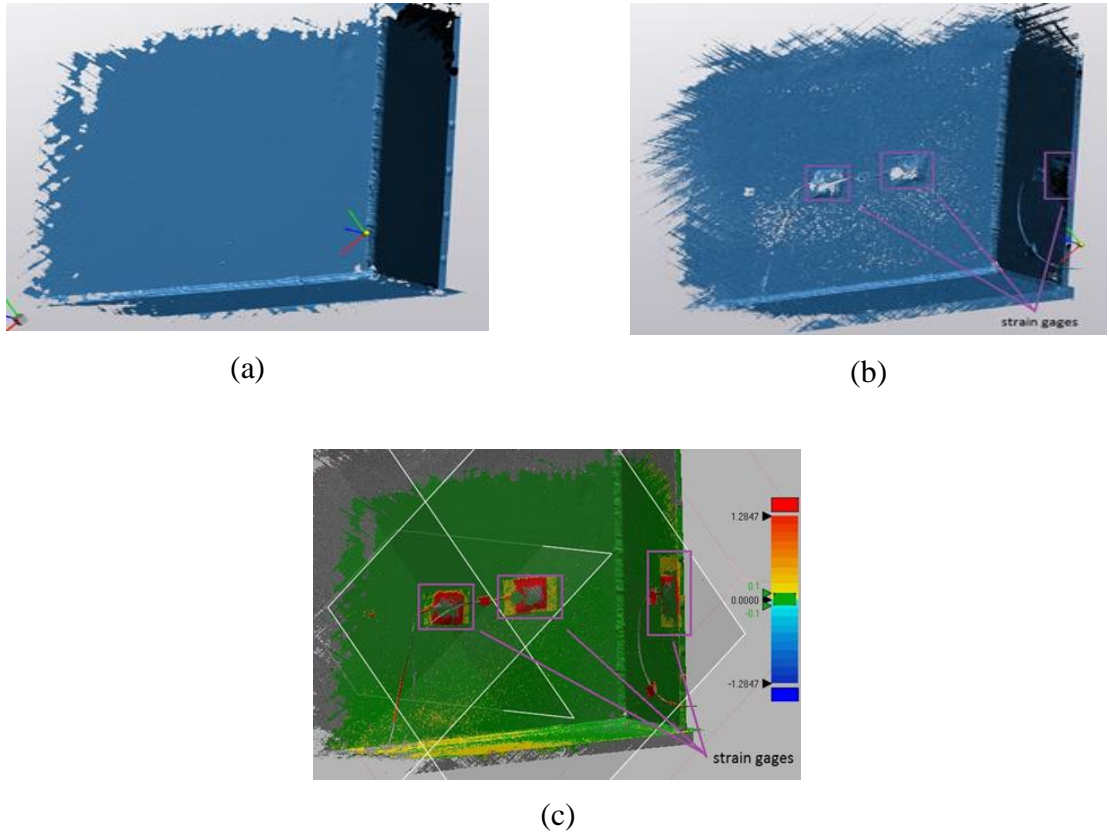

*Figure 6. 3DS of Region 0 (a) scan 1prior to dropping the beam, (b) scan 2 two weeks after dropping the beam, (c) deviation analysis results between Scan 1 and Scan 2 (units in mm)*

## 6. Conclusion

A post-repair evaluation of the Mecklenburg Bridge was performed using a combination of measurements derived from traditional foil resistance strain gauges, full-field digital image correlation, and three-dimensional scanning. Based on the observations from previously repaired beams on this bridge, it was expected that the end region evaluated during this experimental study would exhibit some form of local buckling resulting from uncertainties in the repair; however, following the drop sequence, no observations of local deformation were observed. The results from the measurements did however, prove useful in evaluating the dead load sharing behavior within the bridge. This behavior characteristic is typically assumed in the design and evaluation process and typically difficult to measure. The results from the 3D-DIC measurement demonstrated that for this bridge estimates of dead load sharing behavior used for design, which assumes equivalent load sharing amongst all members, are reasonable approximations based on the inverse approximation derived from the 3D-DIC measurements. This finding is based on a limited study of a single beam location, but provides a framework that could be extended for a more detailed investigation. Results from the study also demonstrated that these 3D full-field measurement approaches are appropriate and suitable for describing 3D shape characteristics with high precision. The cause of the discrepancy between the 3D-DIC measurements and the foil resistance gauges remains uncertain, but the contour results from the 3D-DIC measurements do suggest the potential for local effects as a function of measurement position in the non-ideal structural system.